\input phyzzx
\voffset -0.1in

\def\dL{{\dot L}}
\def\dR{{\dot R}}
\def\dM{{\dot M}}
\def\bP{{\overline P}}
\def\bR{{\overline R}}
\def\dtau{{\dot \tau}}

\centerline{\seventeenrm Mass Quantization of the Schwarzschild Black Hole}
\vskip 0.75in
\centerline{Cenalo Vaz\footnote{1}{Email: cvaz@haar.pha.jhu.edu}
\footnote{2}{On leave of absence from the Universidade do Algarve, Faro, 
Portugal}}
\centerline{\it Department of Physics}
\centerline{\it The Johns Hopkins University}
\centerline{\it Baltimore, MD 21218}
\vskip 0.1in
\centerline{and}
\vskip 0.1in
\centerline{Louis Witten\footnote{3}{Email: witten@physunc.phy.uc.edu}}
\centerline{\it Department of Physics}
\centerline{\it The University of Cincinnati}
\centerline{\it Cincinnati, OH 45221-0011}
\vskip 0.75in

\centerline{\caps Abstract}
{\Tenpoint

\noindent We examine the Wheeler-DeWitt equation for a static, eternal 
Schwarzschild black hole in Kucha\v r-Brown variables and obtain its 
energy eigenstates. Consistent solutions vanish in the exterior of 
the Kruskal manifold and are non-vanishing only in the interior. 
The system is reminiscent of a particle in a box. States 
of definite parity avoid the singular geometry by vanishing at the origin. 
These definite parity states admit a discrete energy spectrum, depending 
on one quantum number which determines the Arnowitt-Deser-Misner (ADM) 
mass of  the black hole according to a relation conjectured long ago by 
Bekenstein, $M\sim\sqrt{n}M_p$. If attention is restricted only to these 
quantized energy states, a black hole is described not only by its mass 
but also by its parity. States of indefinite parity do not admit a quantized 
mass spectrum.
}
\vfill\eject

General dimensional arguments seem to suggest that true quantum gravity 
effects should become manifest only at the Planck scale, $M_p \sim 10^{19}$ 
Gev, $l_p \sim 10^{-35}$ m. Probing such distance scales directly is well beyond 
the realm of experimental possibility, but one may well ask whether or not 
there are non-perturbative effects that manifest themselves at more reasonable 
energy scales, or equivalently at distance scales more accessible to the 
laboratory.

Where would events occur which could have observational implications? 
Probably, the best candidates would involve a system of collapsing matter at 
the very end stages of collapse. According to classical general relativity, a 
very massive star will eventually undergo continuous collapse until it finally 
forms a singularity. The precise nature of the singularity, for example, whether 
it is covered or naked, is not known,${}^{1}$ but there is general agreement 
that naked singularities are pathological and the end state will be a covered 
singularity or a black hole. This expectation has led to a remarkable amount of 
research over the past several decades on the physics of black holes which, in 
turn, has resulted in the unfolding of a long list of interesting properties. 
Perhaps more importantly, black hole physics has forced us to take a harder 
look at some very fundamental, and yet unanswered, questions of principle. Thus, 
(a) is the Hawking radiation${}^{2}$ truly thermal and, if so, why? (b) why 
is the entropy of a black hole proportional to its area?${}^{3,4,5}$ (c) what is 
the end state of an evaporating black hole? (d) what happens to all the 
information that disappeared down the hole? All of these questions are 
related and any attempt to answer them satisfactorily must involve a full and 
consistent quantization of the black hole geometry. 

In this letter we apply midi-superspace techniques${}^{6}$ to study the quantum 
states of an eternal black hole. We conclude that a normalizable wave function must 
vanish everywhere in the exterior of the Kruskal manifold, having support only 
in its interior. They are not {\it a priori} required to have definite parity, but 
the definite parity eigenstates depend on one quantum number and the system 
is reminiscent of a simple particle in a box. States of definite parity do not 
support the singular geometry and, if attention is restricted to them, then a
black hole must be defined both by its mass and parity.  The single quantum number 
determines the energy of the black hole, so the ADM mass is quantized in units 
of the Planck mass according to a formula conjectured and employed by 
Bekenstein,${}^{4,7}$ $M \sim \sqrt{n} M_p$.  The result implies that 
Hawking radiation is {\it not} thermal and has been used to derive the area law of 
black hole thermodynamics.${}^{4}$ In what follows, we will take $\hbar = c = G = 1$.

We begin with the general problem of the collapse of pressureless dust, described 
by an action of the form
$$S~~ =~~ S_g~  +~  S_m~~ =~~ -~ {1 \over {16\pi}} \int d^4 x \sqrt{-g} 
{\cal R}~~ -~~  \int d^4 x \sqrt{-g} \epsilon(x) \left[g_{\alpha \beta} U^\alpha 
U^\beta + 1\right]\eqno(1)$$
where ${\cal R}$ is the scalar curvature, $\epsilon(x)$ is the density of 
the collapsing matter in its proper frame and $U^\alpha$ is the dust velocity.
Although the above action describes the collapse of inhomogeneous, pressureless 
dust in general, we will be concerned with the eternal Schwarzschild black hole 
in this letter. To recover the Schwarzschild vacuum, we will require (at a 
later stage) that the Schwarzschild mass function, $M$, be constant. The dust 
will then become tenuous, playing only the role of a timekeeper according to 
a scheme first introduced by Brown and Kucha\v r.${}^{8}$
 
The action (1) can be cast in canonical form by employing the usual 
ADM${}^{9}$ 3+1 dimensional split of a spherically symmetric spacetime
$$ds^2~~ =~~ N^2 dt^2~ -~ L^2 (dr - N^r dt)^2~ -~ R^2 d\Omega^2, \eqno(2)$$
where $N=N(t,r)$ and $N^r=N^r(t,r)$ are the lapse and shift functions, 
$\epsilon=\epsilon(t,r)$, $L=L(t,r)$, and $R=R(t,r)$ is the physical radius
or curvature coordinate. As emphasized by Regge and Teitelboim${}^{10}$, 
it is important to make no specific gauge choices at this stage of the 
problem; otherwise, essential elements of the canonical structure are 
eliminated. Using (2), the gravitational part of the action together with 
the dust can be cast into the form
$$\eqalign{S~~ =~~ \int dt dr &\left[ P_L \dL~  +~  P_R \dR~  +~ P_\tau \dtau~  
-~  N H~  -~  N^r H_r \right]\cr &+~~ surface~ terms,\cr} \eqno(3)$$
where we have introduced, following Brown and Kucha\v r,${}^{8}$ the dust 
proper time variable, $\tau$, which in general will serve as extrinsic time. 
$P_L$ and $P_R$ are the momenta conjugate to $L$ and $R$ respectively, and 
the super hamiltonian, $H$, and super momentum, $H_r$, are respectively given 
by
$$\eqalign{H~~ =~~ -~ \left[{{P_L P_R} \over R}~  -~  {{L P_L^2}
\over {2R^2}}\right]~  &+~ \left[ -~ {L \over 2}~  -~  {{B'}^2 \over {2L}}~  +~  
\left({{BB'} \over L}\right)'\right]\cr  &+~ P_\tau\sqrt{1~ +   {\tau'}^2/L^2}
\cr}\eqno(4)$$
and 
$$H_r~~ =~~ R' P_R~  -~  L P_L'~  +~  \tau' P_\tau,\eqno(5)$$
where the prime denotes a derivative with respect to the ADM label coordinate $r$. 
The constraints in the above form do not ``decouple'' and are very difficult 
to resolve as they stand. From the general system in (2), Kucha\v r${}^{11}$ 
showed how one can pass by a canonical transformation to a new canonical chart with 
coordinates $M$ and $R$ together with their conjugate momenta, $P_M$ and $\bP_R$, 
where $M$ is the Schwarzschild ``mass'' and $R$ is the curvature coordinate. 
In this system the constraints are greatly simplified and the phase space 
variables have immediate physical significance. The canonical transformation is 
well-defined as long as the metric obeys standard fall-off conditions${}^{11}$ and, 
as long as these fall-off conditions are obeyed, the surface action can be 
recast in the form
$$surface~ terms~~ =~~ \int dt \left[ \pi_+ \dtau_+~  +~  \pi_- \dtau_-~  -~  
N_+ C_+~  -~  N_- C_- \right],\eqno(6)$$
where $\tau_\pm$ are the proper times measured on the parametrization clocks 
at right (left) infinity. The constraints $C_\pm = \pm \pi_\pm + M_\pm$ 
identify their conjugate momenta as the mass at right (left) infinity. 
In terms of the new variables the entire action, along with the surface term is
$$S~~ =~~ \int dt \int dr \left[\bP_M \dM~  +~  \bP_R \dR~  +~  \bP_\tau \dtau~  
-~  NH~  -~  N^r H_r\right],\eqno(7)$$
where $\bP_M = P_M - \tau'$, $\bP_\tau = P_\tau + M'$ and $P_M$ and $P_\tau$ are
the original Brown-Kucha\v r variables. The transformations leading to these 
variables may be found in refs.[8,11]. In passing to the transformed momenta, 
$\bP_\tau$ and $\bP_M$, we have made a canonical transformation generated by 
$M\tau'$, which effectively absorbs the surface terms. (We have implicitly 
fixed the dust proper time to coincide at infinity with the parametrization 
clocks.) 

The super hamiltonian and super momentum constraints become
$$\eqalign{H~~ =~~ -~  &\left[{{F^{-1} M' R' + F \bP_R (\bP_M + \tau')}
\over {L}}\right]\cr &+~~  (\bP_\tau - M') \sqrt{1 + {\tau'}^2
/L^2}~~ =~~ 0\cr}\eqno(8)$$
and
$$H_r~~ =~~ M'\bP_M~  +~  R'\bP_R~  +~  \tau'\bP_\tau~~ =~~ 0, \eqno(9)$$
where we have used
$$L^2~~ =~~ F^{-1} R'^2~ -~ F(\bP_M + \tau')^2 \eqno(10)$$
and $F~~ =~~ 1-2M/R$. $L^2$, being the component $g_{rr}$ of the spherically 
symmetric metric in (2), must be positive definite everywhere. $F$ is positive 
in the exterior (Schwarzschild) region and negative in the interior and this will 
play an important role in the consistency conditions that follow.  By direct 
computation of Poisson brackets, it is easy to determine the ``velocities''  in 
terms of the conjugate momenta from the above expressions and they are 
$$\eqalign{\dtau~~ &=~~ N\sqrt{1+{\tau'}^2/L^2}~  +~  N^r \tau',\cr \dR~~ &=~~
- {{NF(\bP_M+\tau')}\over L}~  +~  N^r R',\cr \dM~~ &=~~ {{NR'\tau' (\bP_\tau-M')}
\over {L^3}}~  +~  N^r M'.\cr}\eqno(11)$$
To proceed with the quantization program, one must raise the canonical variables 
to operator status and consider the constraints in (7) as operator constraints 
on the wave functional $\Psi(\tau,R,M)$, {\it i.e.,} one sets
$${\hat H} \Psi~~ =~~ 0~~ =~~ {\hat H_r} \Psi.\eqno(12)$$
The second constraint enforces spatial diffeomorphism invariance of 
the wave functional on hypersurfaces orthogonal to the dust proper time. It simply 
says that $\Psi' = 0$, where the derivative is with respect to the label coordinate 
$r$. It is convenient to use the super momentum constraint in (9) to eliminate 
$\bP_M$ in the expression for the super hamiltonian in (8). The super hamiltonian 
constraint turns into
$$(\bP_\tau-M')^2~  +~  F \bP_R^2~  -~ {{M'}^2 \over F}~~ =~~ 0.\eqno(13)$$
We will now specialize to the black hole by requiring $M' = 0$ so that only the 
homogeneous mode of $M(t,r)$ survives. The resulting equation decouples and is
hyperbolic in the region $R < 2M$ (the interior of the Kruskal manifold) but 
elliptic in the region $R>2M$ (the exterior). This is because the quantity 
$F$ is negative in the interior, but positive in the exterior -- so the ``kinetic
energy'' there has the ``wrong'' sign. The two distinct solutions must 
agree on the boundary. States of the quantum theory are described by functions of 
the configuration space variables, $\tau, R$ and $M$.

To obtain the wave equation (12), we replace $\bP_\tau = i\nabla_\tau$ 
and $\bP_R = - i \nabla_R$ in (13) with the result
$$\nabla^2 \Psi~~ =~~ G^{ab} \nabla_a \nabla_b \Psi~~ =~~ {\tilde H} \Psi~~ =~~ 0, 
\eqno(14)$$
where $G_{ab}$ is the field space metric, $G_{ab} = {\rm diag}(1,1/F)$ and 
$\nabla_a$ is the covariant derivative with respect to this metric.  Eq. (14) is 
a massless ``Klein-Gordon'' like equation. We will consider only its positive energy 
solutions, $\Psi(\tau,R,M) = e^{-iE\tau} \psi(R,M)$, beginning with the exterior. 

The configuration space metric, $G_{ab} = {\rm diag}(1, 1/F)$ ($F>0$), defines a 
natural measure on the Hilbert space. Because it is flat, it is convenient to 
transform to the coordinate 
$$R_*~~ =~~ \int dR \sqrt{R \over {R-2M}}~~ =~~ \sqrt{R(R-2M)}~ +~ M \ln
[R - M + \sqrt{R(R-2M)}]\eqno(15)$$
and use $R_*$ instead of the original curvature coordinate, $R$, above. Clearly, 
$R_*~ \epsilon~ (M\ln M, \infty)$  and the wave equation, $\partial_\tau^2 \Psi = - 
\partial_*^2 \Psi$, describes the quantum theory whose  Hilbert space is ${\cal H} 
:= {\cal L}^2({\bf R}, dR_*)$ with inner product
$$\langle \Psi_1,\Psi_2\rangle~~ =~~ \int_{M\ln M}^\infty dR_* \Psi_1^\dagger 
\Psi_2~ .\eqno(16)$$
The positive energy solution that is well behaved in the entire range of $R_*$ 
has the form 
$$\Psi_{out}(R,M,\tau)~~ =~~ b(M) e^{-iE(\tau-iR_*)},\eqno(17)$$
where $b(M)$ is an arbitrary function on $M$. We will now argue that$\Psi_{out}$ 
is identically zero. Spatial diffeomorphism invariance of  $\Psi_{out}$ on 
the hypersurface orthogonal to $\tau$ implies that $E(\tau' - iR_*')\Psi =0$. This 
is met by $\tau' = 0 = R'$, by $E=0$, or by $b(M)=0$. The condition 
$\tau' = 0 = R'$ is unacceptable because the positivity of $L^2$ precludes 
$R'=0$ in the exterior, though $\tau'$ can be vanishing. One could take $\Psi = 
b(M)$ as a consistent exterior solution and this was originally proposed by 
Kucha\v r${}^{11}$. Then $E=0$ and $M$ is automatically constant from (11) 
but, because $M$ is a constant and the field space is not compact in this 
region, this solution would not be normalizable. Furthermore, $E = 0$ 
would represent an uninteresting zero total energy solution, so we take 
$b(M) = 0$ in the exterior, {\it i.e.,} $\Psi_{out} = 0$. 

Next let us consider the solution in the internal region, $R<2M$ ($F<0$). Here 
the equation is hyperbolic and it is convenient to transform to the coordinate 
$\bR_*$ defined by
$$\bR_*~~ =~~ -~ \sqrt{R(2M-R)}~ +~ M \tan^{-1}\left[{{R-M} \over {\sqrt{R(2M-R)}}}
\right]~. \eqno(18)$$
The new coordinate lies in the range $(-{{\pi M} \over 2},+{{\pi M} \over 2})$ and 
the wave equation, $\partial_\tau^2 \Psi = \partial_*^2 \Psi$ now defines the 
quantum theory whose Hilbert space is ${\cal H} := {\cal L}^2({\bf R}, 
d\bR_*)$ with inner product 
$$\langle \Psi_1,\Psi_2\rangle~~ =~~ \int_{-{{\pi M}\over 2}}^{+{{\pi M}\over 2}} 
d\bR_* \Psi_1^\dagger\Psi_2~ . \eqno(19)$$
The general (positive energy) solution is
$$\Psi_{in}~~ =~~ c_+(M) e^{-iE(\tau + \bR_*)}~~ +~~  c_-(M) e^{-iE(\tau - \bR_*)} 
\eqno(20)$$
where $c_\pm$ are functions only of $M$. Again we must impose the super momentum 
constraint, which reads
$$(\tau'+R') c_+(M) e^{-iE(\tau+\bR_*)}~~ +~~ (\tau'-R') c_-(M) e^{-iE(\tau-
\bR_*)}~~ =~~ 0, \eqno(21)$$
assuming $E>0$. A consistent and physically meaningful solution to this equation 
is $\tau' = \bR'_* = 0$. Returning to (11), we see that the choice implies that 
$\dtau = N$ and $\dM=0$. Setting $N=1$, the dust proper time turns into the 
asymptotic Minkowski time and the energy, $E$, should be associated with the 
ADM mass of the black hole.

This solution must match the solution in the exterior at $R=2M$. Matching gives
$$c_-(M)~~ =~~ -~  c_+(M)e^{-iEM\pi},\eqno(22)$$
so that
$$\Psi_{in}~~ =~~ c_+(M) \left[ e^{-iE(\tau+\bR_*)}~~  -~~ e^{-iEM\pi} 
e^{-iE(\tau -\bR_*)}\right]~ .\eqno(23)$$
Because the parity operator, $\bR_* \rightarrow -\bR_*$, commutes with the 
``Hamiltonian'' operator, ${\tilde H}$, states of definite parity are guaranteed 
to remain so for all times. The definite parity eigenstates, apart from 
obeying the symmetries of ${\tilde H}$ and the domain of $\bR_*$, vanish at 
$R=0$ ($\bR_* = -{{\pi M} \over 2}$). Therefore they provide no support for the 
classical singular geometry. These definite parity eigenstates exhibit a 
discrete energy spectrum and are given by 
$$\eqalign{\Psi_{in}^{(+)}~~ &=~~ {1 \over {\sqrt{\pi M}}} e^{-iE\tau} 
\cos E\bR_*~~~~~~~~~~ EM~ =~ (2n+1)~ ,\cr \Psi_{in}^{(-)}~~ &=~~ {1 \over 
{\sqrt{\pi M}}} e^{-iE\tau} \sin E\bR_*~~~~~~~~~~ EM~ =~ 2n~ ,\cr}
\eqno(24)$$
where we take $n~ \epsilon~ {\bf N} \cup \{0\}$ to maintain the positivity of the 
total energy, that is, to agree with the classical positive energy theorems. If we 
identify the total energy with the ADM mass of the black hole, then the ADM mass 
is quantized in units of the Planck mass, 
$$M = \sqrt{n} M_p,~~~~~~~~~~~~~~~ n~ \epsilon~ {\bf N} \eqno(25)$$ 
as proposed in the introduction.  Restricting attention to 
states of definite parity seems very natural for the reasons mentioned above. It raises 
an intriguing possibility: that quantum black holes are determined by their parity 
as well as by their mass.

Considerations involving the quantization of the angular momentum and charge of 
extremal and non-extremal Reissner-Nordstrom and Kerr-Newman black holes led 
Bekenstein to propose, long ago, that the horizon area of a black hole was 
quantized in integer multiples of a fundamental area, presumably of the order 
of $l_p^2$. Because the area of the horizon is proportional to the square of 
the mass of a black hole, Bekenstein's original proposal can be viewed as a 
proposal for the spectrum of the (quantum) mass operator and coincides exactly 
with our formula (24). The consequences of this mass spectrum have been discussed 
extensively by Bekenstein and Mukhanov.${}^{12}$ Many attempts, ranging from 
quantum membrane and string theory approaches${}^{13}$, to canonical quantum 
gravity treatments${}^{14}$ of collapsing matter and vacuum spacetimes, have 
been made to obtain Bekenstein's formula from first principles. These attempts 
have had varying results which depend strongly on the simplifying assumptions 
made either in the model itself or in the steps leading to the Hamiltonian 
reduction. Some obtain equally spaced area levels and others do not.

We have used the Hamiltonian reduction of spherical geometries by Kucha\v r${}^{11}$ 
and the coupling to incoherent dust by Brown and Kucha\v r${}^{8}$ to quantize the 
Schwarzschild black hole. We have only required homogeneity and the positivity 
of the black hole mass. Thus the dust was made tenuous and clocks attached 
to the dust particles served to identify the time foliation. Time evolution appeared 
naturally in the formalism. An appropriate choice of the lapse 
function fixed the dust proper time to coincide with the proper time of a static 
asymptotic observer and the total energy was identified with the ADM mass of the 
hole. A quantum black hole behaves very much like a particle in a box. 
The wave function vanishes in the exterior, therefore the dynamics of the system 
takes place in the interior of the Kruskal manifold. This reflects the fact that 
the exterior region is a vacuum for the asymptotic observer. In the interior, 
normalized solutions of definite parity are quantized in half-integer units. We know 
of no reason to select only states of definite parity, other than the fact that 
parity is a discrete symmetry of the system and that these states exhibit a node 
at the classical singularity. This means that they do not support the singular 
geometry. 

In this letter we have confined our study to the case of the eternal Schwarzschild 
black hole though we foresee no obstacle to treating charged and rotating black holes.
Eq. (13) describes the general problem of the collapse of incoherent dust. Classically, 
models of pressureless dust collapse lead both to covered and naked singularities. 
Semi-classical considerations indicate that the Hawking radiation is vastly different in 
the two cases.${}^{15}$ The natural next step is to examine Hawking radiation in this 
context, but additional degrees of freedom that carry the Hawking radiation must then 
be introduced.
\vskip 0.5in

{\noindent {\bf ACKNOWLEDGEMENTS}}

\noindent We acknowledge the partial support of the  Funda\c{c}\~ao
para a Ci\^encia e Tecnologia (FCT) Portugal, under contract number 
CERN/S/FAE/1172/97 and the partial support of NATO, under contract number 
CRG 920096. L. W. acknowledges the partial support of the U. S. Department 
of Energy under contract number DOE-FG02-84ER40153.
\vskip 0.5in

\noindent{\bf REFERENCES}

{\item{1.}}For recent reviews, see: R. Penrose in {\it Black Holes and Relativistic 
Stars} ed. R.M. Wald (Chicago University Press, 1998); R. M. Wald, gr-qc/9710068; 
C. J. S. Clarke, Classical and Quantum Gravity {\bf 10}, {1375} (1993); P. S. 
Joshi in {\it Singularities, Black Holes and Cosmic Censorship} ed. P. S. Joshi
(IUCAA, Pune, 1997), gr-qc/9702036; P. S. Joshi, {\it Global Aspects in
Gravitation and Cosmology} (Oxford, 1993); T. P. Singh in {\it Classical and
Quantum Aspects of Gravitation and Cosmology}, ed. G. Date and B. R. Iyer
(Inst. of Math. Sc., Madras, 1996), gr-qc/9606016.

{\item{2.}}S. W. Hawking, Comm. Math. Phys. {\bf 43} (1975) 199.

{\item{3.}}S. W. Hawking, Phys. Rev. Lett. {\bf 26} (1971) 1344; 

{\item{4.}}J. D. Bekenstein, Lett. Nuovo Cimento {\bf 11} (1974) 467.

{\item{5.}}V. Mukhanov, JETP Letts. {\bf 44} (1986) 63.

{\item{6.}}B. S. DeWitt, Phys. Rev. {\bf 160} (1967) 1113; K. V. Kucha\v r, Phys. 
Rev. {\bf D4} (1971) 955; in {\it Quantum Gravity II: Second Oxford Symposium},
ed. C. J. Isham, R. Penrose and W. Sciama (Clarendon, Oxford, 1981).

{\item{7.}}J. D. Bekenstein, ``Black Holes: Classical Properties, Thermodynamics 
and Heuristic Quantization'', in the IX Brazilian School on Cosmology and 
Gravitation, Rio de Janeiro 7-8/98, gr-qc/9808028; ``Quantum Black Holes as 
Atoms'', in the VIII Marcel Grossmann Meeting on General Relativity, 
Jerusalem, June 1997, gr-qc/9710076; Phys. Lett. {\bf B360} (1995) 7; Phys. 
Rev. Lett. {\bf 70} (1993) 3680.

{\item{8.}}J. D. Brown, K. V. Kucha\v r, Phys. Rev. {\bf D51} (1995) 5600.

{\item{9.}}R. Arnowitt, S. Deser and C.W. Misner, in {\it Gravitation: An 
Introduction to Current Research}, ed. Louis Witten (Wiley, New York, (1962).

{\item{10.}}T. Regge and C. Teitelboim, Ann. Phys. {\bf 88} (1974) 286.

{\item{11.}}K. V. Kucha\v r, Phys. Rev. {\bf D50} (1994) 3961.

{\item{12.}}J. D. Bekenstein and V. F. Mukhanov in {\it Sixth Moscow Quantum 
Gravity Seminar}, ed. V. A. Berezin, V. A. Rubakov and D. V. Semikoz 
(World Publishing, Singapore, 1997); V. Mukhanov, in {\it Complexity, Entropy 
and the Physics of Information: SFI Studies in the Sciences of Complexity}, 
Vol III, ed. W. H. Zurek (Addison-Wesley, New York, 1990).

{\item{13.}}A. W. Peet, Class. Quant. Grav. {\bf 15} (1998) 3291; K. Sfetsos, 
K. Skenderis, Nucl. Phys. {\bf B517} (1998) 179; A. Strominger and C. Vafa, 
Phys. Lett. {\bf B379} (1996) 99;  C. O. Lousto, Phys. Rev. {\bf D51} (1995) 
1733; M. Maggiore, Nucl. Phys. {\bf B429}  (1994) 205;  Ya. I. Kogan, JETP
Lett. {\bf 44} (1986) 267.

{\item{14.}}V. A. Berezin, A. M. Boyarsky and A. Yu. Neronov, Phys. Rev. 
{\bf D57} (1998) 1118; S. Hod, Phys. Rev. Lett. {\bf 81}  (1998) 4293; V. A. 
Berezin, Phys. Rev. {\bf D55} (1997) 2139; 
K. V. Krasnov, Phys. Rev. {\bf D55} (1997) 3505; A. Ashtekar, J. Lewandowski, 
Class. Quant. Grav. {\bf 14} (1997) A55; J. Louko and J. M\"akel\"a, Phys. 
Rev. {\bf D54} (1996) 4982; H. A. Kastrup, Phys. Lett. {\bf B385} (1996) 75;
Y. Peleg, Phys. Lett. {\bf B356} (1995) 462; 

{\item{15.}}Sukratu Barve, T.P. Singh, Cenalo Vaz and Louis Witten, Phys. Rev. 
{\bf D58} (1998) 104018; Nucl. Phys. {\bf B532} (1998) 361; Cenalo Vaz and 
Louis Witten, gr-qc/9804001, Phys. Lett. {\bf B} {\it in press}.

\bye